\documentclass{aastex6}
\usepackage{graphicx}

\begin{document}

\def\apj{{ApJ}}
\def\mnras{{MNRAS}}
\def\aa{{A\&A}}
\def\Nature{{Nature}}
\def\GCN{{GCN Circ}}
\def\PRD{{Phys. Rev. D}}
\def\PRL{{Phys. Rev. Lett}}
\def\etal{{\it et al.~}}

\newcommand{\xmm}{XMMU~J183245$-$0921539}
\newcommand{\mass}{2MASS~J18324516$-$0921545}
\newcommand{\hess}{HESS~J1832$-$093}
\newcommand{\gr}{$\gamma$-ray}
\newcommand{\grs}{$\gamma$-rays}
\newcommand{\gcn}{GCN Circ.}

\shorttitle{HESS J1832-093 as a \gr~binary}
\shortauthors{Tam et al.}

\title{A multi-wavelength study of the \gr~binary candidate HESS~J1832-093}

\author{
Pak-Hin Thomas Tam\altaffilmark{1}, K. K. Lee\altaffilmark{2}, Yudong Cui\altaffilmark{1}, A. K. H. Kong\altaffilmark{3}, K. L. Li\altaffilmark{4}, Vlad Tudor\altaffilmark{5}, Xinbo He\altaffilmark{1}, C.~P. Hu\altaffilmark{6,7}, Partha S. Pal\altaffilmark{1}}
\affil{$^1$ School of Physics and Astronomy, Sun Yat-sen University, Zhuhai 519082, China\\
$^2$ Department of Physics, The Chinese University of Hong Kong, Shatin, N.T., Hong Kong SAR, China\\
$^3$ Institute of Astronomy, National Tsing Hua University 101, Section 2. Kuang-Fu Road, Hsinchu, 30013, Taiwan, R.O.C. \\
$^4$ Department of Physics, UNIST, Ulsan 44919, Korea \\
$^5$ International Centre for Radio Astronomy Research, Curtin University, GPO Box U1987, Perth, WA 6845, Australia \\
$^6$ Department of Astronomy, Kyoto University, Kitashirakawa-Oiwake-cho, Sakyo-ku, Kyoto 606-8502, Japan \\
$^7$ JSPS International Research Fellow}
\email{tanbxuan@mail.sysu.edu.cn}

\begin{abstract}
We investigate the nature of the unidentified very-high-energy (VHE) \gr~object, HESS~J1832-093, in a multi-wavelength context. Based on X-ray variability and spectral index ($\Gamma_X\sim\,1.5$), and its broad-band spectrum (which {\bf was} remarkably similar to HESS J0632+057, a confirmed \gr~binary), HESS~J1832-093 has been considered to be a strong \gr~binary candidate {\bf in previous works}. In this work, we provide further evidence for this scenario. We obtained a spectrum of its IR counterpart using Gemini/{\bf Flamingo}, finding absorption lines that are usually seen in massive stars, in particular O stars. We also obtained a rather steep ATCA spectrum ($\alpha=-1.18^{+1.04}_{-0.88}$) which prefers a \gr~binary over an AGN scenario. Based on spatial-spectral analysis and variability search, we found that 4FGL J1832.9-0913 {\bf is possible to be associated with SNR~G22.7-0.2 rather than with HESS~J1832-093 only.}
\end{abstract}


\section{Introduction}
\gr~binaries are a class of high-mass X-ray binaries which radiates dominantly in the \gr~energy band in the $\nu F_\nu$ representation~\citep{Dubus13}. Composing of a neutron star or a black hole, and a O/Be companion, \gr~binaries consist of only {\bf several well established members until 2019}: PSR~B1259$-$63, LS~I~61~303, LS~5039, HESS~J0632-057, {\bf 1FGL~J1018-5658 \citep{Dubus15,hess0632_gev}, PSR~J2032$+$4127 \citep{lyne15,ho17,ab18}, 4FGL J1405.1-6119 \citep{Gorbet2019}, and a point source ``P3'' in the Large Magellanic Cloud (LMC) \citep{Corbet16,HESS_lmc_p3}.} All of them have been detected above 100~GeV by ground-based Cherenkov telescopes {\bf and/or } at 100~MeV to 100~GeV by the Large Area Telescope (LAT) onboard the {\it Fermi} satellite. 

The Cherenkov telescope array, High Energy Stereoscopic System (H.E.S.S.), was used to discover HESS~J1832$-$093~in the vicinity of SNR G22.7$-$0.2. It is seen as a point source by H.E.S.S., and its 0.4$-$5~TeV spectrum is well fit by a single power law with photon index of $\Gamma_\mathrm{TeV}=2.6\pm0.3_\mathrm{stat}\pm0.1_\mathrm{syst}$. Its flux is around 1\% that of the Crab nebula above 1~TeV. The observations were carried out from 2004 to 2011, comprising 67 hours live time, and no variability were found in the data~\citep{hess_1832_discover}. Based on its vicinity with SNR G22.7$-$0.2 and a spatially coincident CO emission, an SNR-molecular cloud interaction scenario was suggested. \citet{hess_1832_discover} also used $>$10~GeV data from the {\it Fermi}/LAT and did not find any significant emission.

 \hess~has been observed with various X-ray instruments since 2008. An X-ray source, \xmm, consistent with the location of HESS~J1832$-$093, was found in a dedicated {\it XMM-Newton} observation in 2011~\citep{hess_1832_discover}. While the X-ray column density of $N_\mathrm{H}\sim10^{23}$~cm$^{-2}$ toward the source is high~\citep{eger16,mori17}, its X-ray emission is at the level of 10$^{12}$~erg~cm$^{-2}$~s$^{-1}$ with a hard photon index of $\sim$1.5. A {\it Chandra} observation in July 2015 (MJD~57209) was used to refine the {\it XMM-Newton} source location and it is fully consistent with an infrared (IR) source, \mass~\citep{eger16}. It is very likely that the IR, X-ray and TeV sources are associated with each other based on the spatial coincidence. The {\it Chandra} observation also constrained the X-ray source size to be $<$0.28 arcsec with 90\% confidence. {\it NuSTAR} detects X-rays up to 30 keV with a $\Gamma_\mathrm{X}=$1.5$\pm$0.1 spectrum, showing particle acceleration typically seen in Galactic \gr~binary systems. No X-ray periodicity from 4ms to 85.7 ks was found in the {\it NuSTAR} data~\citep{mori17}.

\citet{eger16}~reported that {\it Chandra} saw \xmm~at an elevated X-ray flux state (on MJD~57209; about a factor of 6 above other measured X-ray flux), while the X-ray flux at other epochs (obtained by Swift/XRT and {\it XMM-Newton} observations) are consistent with being constant~\citep{eger16}. However, this level of variability is disputed~\citep{mori17}, who instead suggest an X-ray variability at the level of 50\% based on the {\it XMM-Newton} and {\it NuSTAR} flux at two different epochs. If confirmed, such a variability almost certainly rules out scenarios that predict steady emission, including SNR-molecular cloud interaction or a putative pulsar wind nebula. Instead, as discussed in \citet{eger16}, we seem to left with two scenarios: a \gr~binary or an active galactic nucleus (AGN).

There is no known optical counterpart, and it is likely because of high level of absorption revealed by the X-ray data. Spitzer detected the IR counterpart, \mass, up to 8~$\mu$m (from the catalogue; source name SSTGLMC G022.4768-00.1539). Pan-STARRS have a few measurements in the z (=20.32+/-0.135) and y (=19.702+/-0.05) bands.

To further probe the origin of HESS~J1832$-$093, we obtained Gemini-south NIR spectroscopic observations and investigate the {\it Fermi}-LAT data obtained over the first 10.6 years. We also obtained data by ATCA (2016 June 2) and Swift/XRT (up to MJD 58287). Based on the multi-wavelength datasets, we argue that HESS~J1832$-$093 is indeed a \gr~binary. 

\section{Gemini near-infrared observation}

\begin{figure}
    \epsscale{1.}
    \plotone{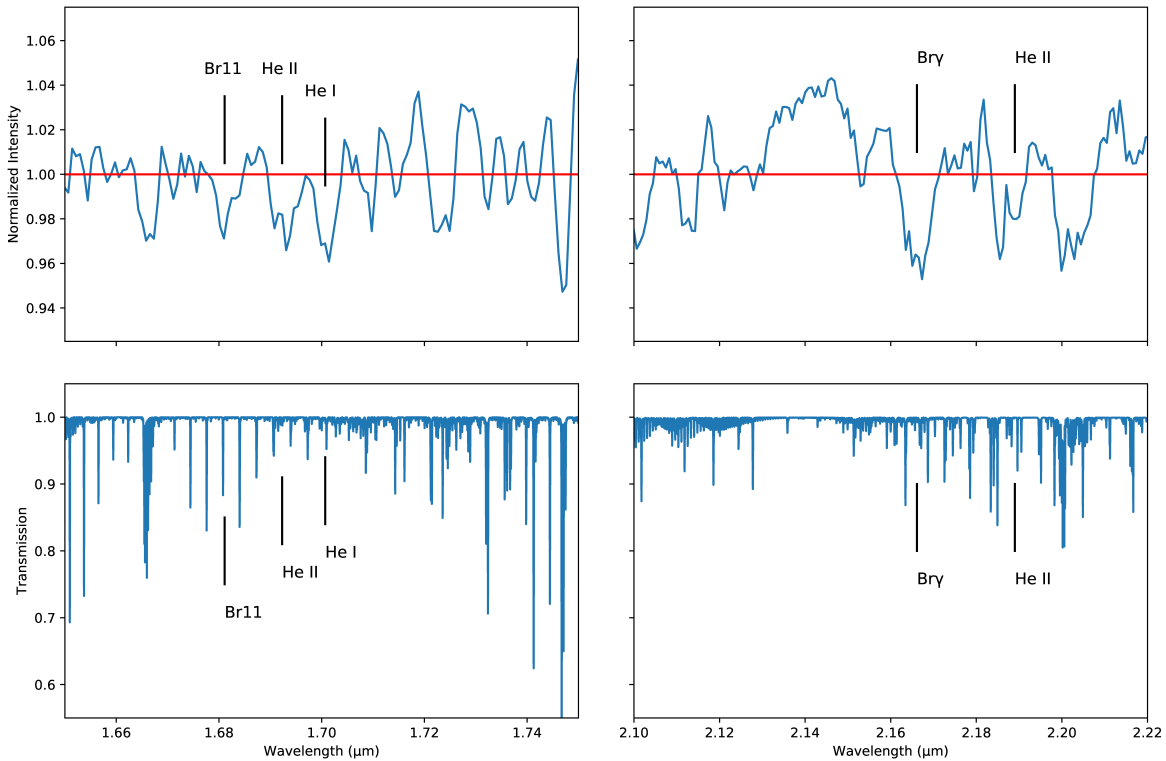}
      \caption{{\it Top panel}: The H-band (left) and K-band (right) spectrum taken with FLAMINGOS-2 at the Gemini-south 
Observatory on June 2, 2016. {\it Bottom panel}: the IR transmission spectra taken from the Gemini website.}
      \label{fig:gemini_spec}
   \end{figure}

We obtained a Fast Turnaround (FT) mode of observation using the FLAMINGOS-2 (long-slit spectrograph) instrument at the Gemini-south Observatory on June 2, 2016. The data reduction was done using the IRAF\footnote{IRAF is distributed by the National Optical Astronomy Observatories.} software. The spectra that we obtained are shown in the upper panel of Fig.~\ref{fig:gemini_spec}. The data quality is not particularly high and we note several strong telluric absorption lines due to water vapor. 

To quantify the signal-to-noise ratio of any stellar lines among the telluric lines,  we define $r$ to be (Area of the absorption lines measured - Area of the telluric absorption lines)/(Area of the telluric absorption lines), and identify four lines with $r>4$, which are shown in {\bf Table~\ref{tab:ew}}, together with their equivalent widths. The region longer than 1.72$\mu$m in H-band is severely contaminated, and the contamination is more severe in H-band than in K-band. To make the situation more explicit, we depict in the bottom panel of Fig.~\ref{fig:gemini_spec} the IR transmission spectra taken from the Gemini website~\footnote{\url{https://www.gemini.edu/sciops/telescopes-and-sites/observing-condition-constraints/ir-transmission-spectra}}. 

   \begin{table}
\begin{center}
\caption{The $r$-value and equivalent widths (EWs) of each identified line feature.
\label{tab:ew}}
\begin{tabular}{ccc}
\hline
Line & $r$ & EW($\AA$) \\
\hline
Br11 (1.6811$\mu$m) & 8.55 & $1.49\pm0.44$ \\
Br$\gamma$ (2.1661$\mu$m) & 7.48 & $3.75\pm0.59$ \\
He I (1.7007$\mu$m) & 6.45 & $1.89\pm0.39$ \\
He II (1.6923$\mu$m) & 4.59 & $1.68\pm0.41$ \\
He II (2.189$\mu$m) & 4.13 & $1.44\pm0.34$ \\
\hline
\end{tabular}
\end{center}
\end{table}
We found evidence of the lines related to Br$\gamma$, Br11, He I, and He II, many of which are seen in massive stars only:
\begin{enumerate}
\item The He II 2.189 μm line is seen in O stars but not B stars~\citep{martins19};
\item He II (1.6923$\mu$m) line appears only in the O-type star~\citep{hanson98, roman-lopes18};
\item The equivalent width of He I (1.7007$\mu$m) line shows that the spectral type should be before O8 or after B2~\citep{blum97};
\item The equivalent width of Br11 (1.6811$\mu$m) line would indicate that the star is an O-B1 star~\citep{roman-lopes18}.
\end{enumerate}
Therefore, combining all the information above, we suggest that the IR counterpart of \xmm~to be an O- or, less likely, early B-type star. The lack of prominent emission line in NIR also rules out a circumstellar disk normally associated with a Be star. Therefore, we conclude that the NIR counterpart is a massive star, and not an AGN. This conclusion is consistent with the photometric considerations (in J- and K-band) reported in~\citet{mori17}.

\section{{\it Fermi}/LAT observations of HESS~J1832$-$093}

The LAT detector is an all-sky monitor at energies from several tens of MeV to more than 300~GeV~\citep{lat_technical}. The $\gamma$-ray data\footnote{provided by the FSSC at \url{http://Fermi.gsfc.nasa.gov/ssc/}} used in this work were obtained using the {\it Fermi}/LAT between 2008 August 4 and 2019 March 13. We used {\it Fermi}tools 1.0.0 to reduce and analyze the data. Pass 8 data with energy 100~MeV to 500~GeV classified as ``source'' events were used. To reduce the contamination from Earth albedo $\gamma$-rays, events with zenith angles greater than 90$^\circ$ were excluded. The instrument response functions ``P8R3\_SOURCE\_V2'' were used.

\subsection{Maximum Likelihood Analysis}
\label{sect:lat_analysis}
We carried out a binned maximum-likelihood analysis (\emph{gtlike}) of a rectangular region of 15$^\circ\times$15$^\circ$ centered on the position of HESS~J1832$-$093. We subtracted the background contribution by including the Galactic diffuse model (gll\_iem\_v07.fits) and the isotropic background (iso\_P8R3\_SOURCE\_V2\_v1.txt) as well as
the 4FGL~\citep{4fgl_paper} sources within 20$^\circ$ away from HESS~J1832$-$093. FL8Y\footnote{\url{https://Fermi.gsfc.nasa.gov/ssc/data/access/lat/fl8y/}} sources were also employed (instead of 4FGL sources) for cross-check purpose. 
In the FL8Y/4FGL catalog, one source, 4FGL J1832.9-0913 (i.e., FL8Y J1832.5-0921), is said to be associated with HESS J1832-093/SNR G022.7-0.2, in accordance to~\citet{hess_1832_discover}. Moreover, both catalog sources locate very close to HESS~J1832$-$093, hence we did not treat FL8Y J1832.5-0921/4FGL J1832.9-0913 as a background source in our analysis. The \gr~pulsars PSR~J1831-0952 and PSR~J1833-1034 are within one degree from \hess. The former \gr~pulsar was reported in \citet{Laffon15_new_psr} and no  timing ephemeris has been given. Considering also that the current available timing ephemeris of the latter is valid until 2013 October only, c.f. \url{http://www.slac.stanford.edu/~kerrm/Fermi\_pulsar\_timing/}, we did not perform pulsar gating in the likelihood analysis. We, however, include 4FGL J1831.5-0935 (associated with PSR~J1831-0952) in the background model as this pulsar is within 0.4 degree from \hess. The recommended spectral model for each source as in the corresponding catalog was used, while we modeled a putative source exactly at the position of HESS~J1832$-$093~with a power law (PL)
\begin{equation}
\frac{dN}{dE} = N_0 \left(\frac{E}{E_0}\right)^{-\Gamma},
\end{equation}
where the normalization $N_0$ and spectral index $\Gamma$ were allowed to vary. The normalization parameter values for the Galactic and isotropic diffuse components, and for the sources within 6$^\circ$ from HESS~J1832$-$093, the normalization $N_0$ and spectral index $\Gamma$ were allowed to vary as well. 

\begin{figure}
    \epsscale{1.}
   \plotone{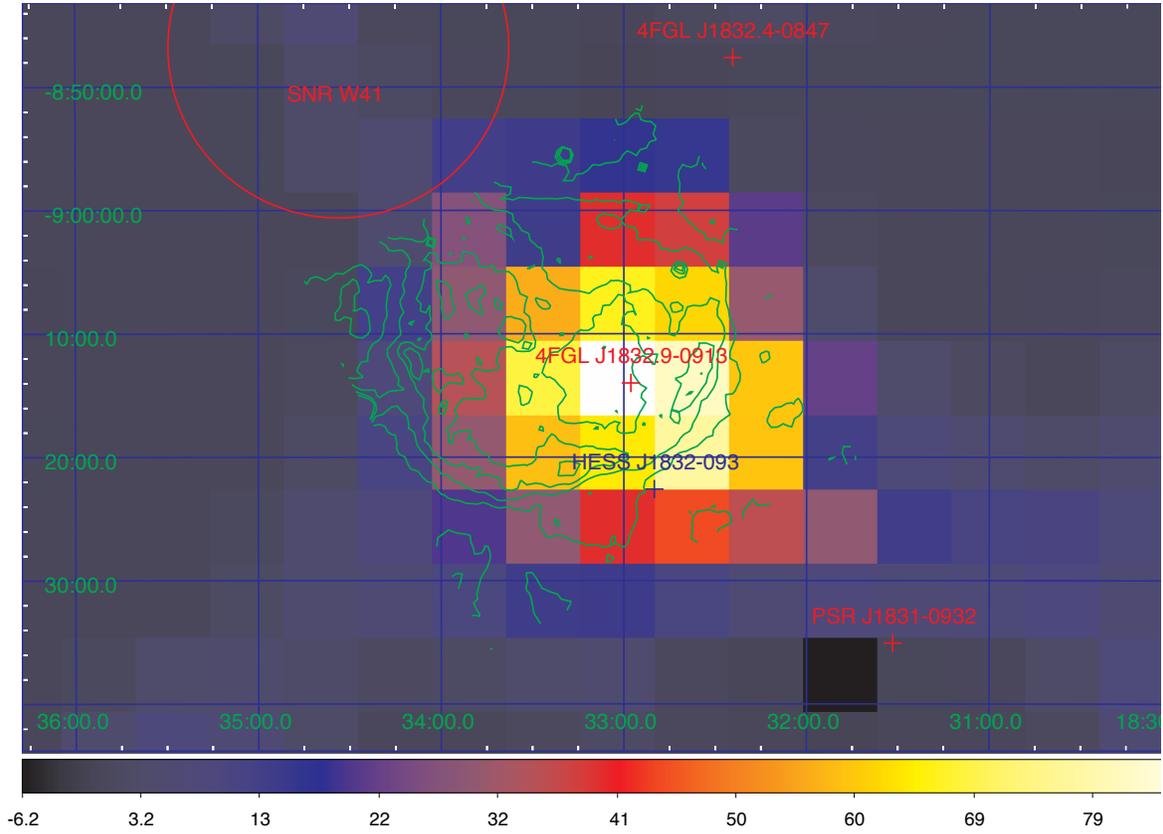}
      \caption{TS map of the 1~deg $\times$ 1~deg region around 4FGL~J1832.9-0913, using 1--500~GeV photons, overlaid in green contours is the 20~cm emission of G022.7-0.2~\citep[from the Multi-Array Galactic Plane Imaging Survey,][]{white05,helfand06}. The x- and y-axis shows the R.A. and Decl., respectively.}
      \label{fig:tsmap}
   \end{figure}

\hess~was detected with TS=95.8 in this analysis. The best-fit power-law index is $2.46\pm0.70$, and the photon flux is (5.10$\pm$1.48)$\times$10$^{-8}$~photons~cm$^{-2}$~s$^{-1}$. We repeated the analysis using photons within narrower energy bands (i.e., 0.1-0.3~GeV, 0.3-1.0~GeV, 1-3~GeV, 3-15~GeV, and 15-300~GeV), and the flux obtained is plotted in the spectral energy distribution (see Fig.~\ref{fig:sed}).
We also attempted to use a power-law with exponential cutoff (PLE), {\bf and the best fit of PLE gives a spectral index of $\Gamma=1.58 \pm 0.29$ and a cutoff energy of $E_\mathrm{c}=650 \pm 140$~MeV. However, PLE shows no statistical improvement.}
To produce the TS map shown in Fig.~\ref{fig:tsmap}, we use 1--500~GeV clean, front-converted photons to avoid low-energy photon contamination from nearby sources.

As the SNR G22.7-0.2 is only 11.5$\arcmin$ from \xmm, and the former can potentially contribute to or even dominate the \grs~that we see (e.g., by SNR-molecular cloud interaction), we proceeded to add a hypothetical point source at the center of the SNR (since there is no a-priori knowledge about which molecular clouds the SNR may be interacting) and performed the maximum-likelihood analysis again. Comparing the three models (\hess~only, G22.7-0.2 only, and both \hess~and G22.7-0.2), the value of -log(likelihood) is the lowest with both sources added (-38552671.38), suggesting that one source among the two dominate the \gr~emission. The value of -log(likelihood) is comparable for fits using \hess~only (-38552668.78) or G22.7-0.2 only (-38552670.5). Therefore there is no strong preference as for which source (\hess~versus G22.7-0.2) is the dominant GeV-emitting source (in other words, to be associated with 4FGL J1832.9-0913) based on this comparison only.

The exact position of 4FGL J1832.9-0913 can be estimated by the pixel in the TS map with the highest TS values, which is R.A.=18$^{\mathrm h}$33$^{\mathrm m}$02$\fs$04, Decl.=$-$09$^\circ$13$\arcmin$43$\farcs$3 (J2000).

\begin{figure}
    \epsscale{0.8}
   \plotone{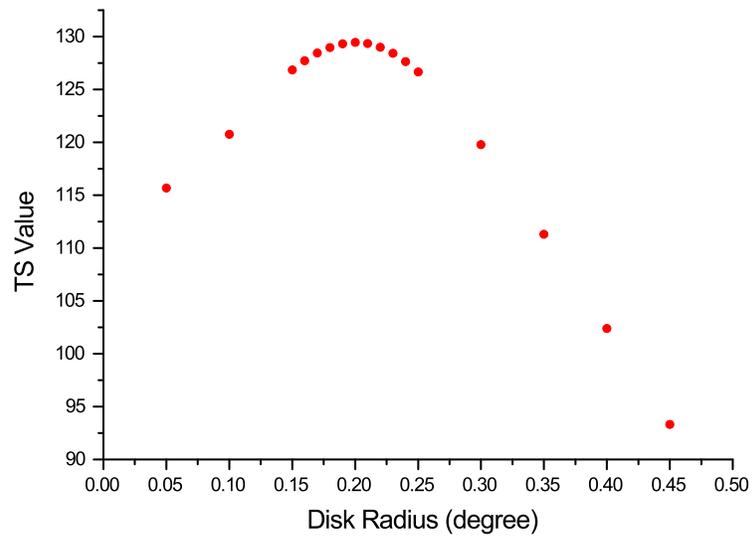}
      \caption{TS value of 4FGL J1832.9-0913 modeled with uniform disks of various radii}
      \label{fig:extension}
   \end{figure}

Another test to help distinguish whether 4FGL J1832.9-0913 originates from \hess~or G22.7-0.2, is to check the possible spatial extension of the source. For this purpose, we insert uniform disks of various radius, from 0.05$^\circ$ to 0.45$^\circ$, centered on the TS peak position shown above. In this analysis, we used spatial bins of 0$\fdg$02 to increase spatial resolution, while allowed the spectral parameters of sources within 3$^\circ$ from HESS~J1832$-$093 to vary.  Fig.~\ref{fig:extension} shows how the TS value of the source varies with increasing radius (which is an indicator of the possible extension). It can be seen that the highest TS value occur for a radius of 0.2$^\circ$ (or 12$\arcmin$), and may suggest possible extension for 4FGL J1832.9-0913.

\subsection{Periodicity and Variability Search}
One distinguishing feature of \gr~binaries are their orbital periodicity. We tried box-fitting algorithm~\citep{box_fitting} and did not find any significant periodicity for time scales from 10$^{-1}$s to 10$^3$s. We also created a \gr~light curve using a bin size of 1~day and search for periodicity with Lomb-Scargle algorithm~\citep{lomb76,scargle82}. We further use REDFIT algorithm to estimate the red noise level~\citep{schulz02}. Only a peak located at $\sim$53.4 day, which is the precession period of the {\it Fermi} orbit, is clearly above the red noise level. 

To probe any possible variability in GeV \grs, we produced light curves using the likelihood analysis for time bins of sizes 173.5 days, which is a balance between small photon statistics and the desire to probe variability on time scales shorter than about a year. The background source model used in the previous section is also used here. 
We do not see any significant change in flux nor spectral index in different time bins. A 88 day-binned light curve is also produced and again no strong variability can be seen. 

To formally assess whether \hess~varies at GeV energies, we follow the same treatment for variability as in 2FGL~\citep{lat_2nd_cat}, as well as in the 3FGL~\citep{3fgl_cat} catalog, by comparing the difference of the likelihood in the null hypothesis (the flux of \hess~is stable over the full time range) and that in the alternative hypothesis (the source flux is allowed to vary). The variability index thus obtained is 26.9, which for 21 degrees of freedom corresponds to a variability confidence level just below 90\% (which would have a variability index 29.6). Therefore, we do not detect any significant variability. 

\section{ATCA RADIO OBSERVATIONS}

We conducted radio continuum observations with ATCA under project CX359, when the array was in 1.5B configuration. These were carried out on 2 June 2016, 14:56 to 19:04 UT, for a total of 3.6h on-source. The Director's discretionary Time (DDT) observations were prompted by the large X-ray flux variability reported by~\citet{eger16}. We simultaneously observed two frequency windows, centred at 5.5 and 9 GHz, each with a bandwidth of 2 GHz. We used the source PKS 1934-638 as a bandpass and flux calibrator, and 1829$-$106 as a phase calibrator.

The data were flagged and calibrated in Miriad v1.5~\citep{Sault95} using standard procedures. We imaged the data in CASA 4.5~\citep{McMullin07} using Briggs weighting of robustness 1, which balances sensitivity and spatial resolution.

Resolutions of 44.7''x1.5'' at 5.5 GHz and 25.7''x1.2'' at 9 GHz were achieved. We detected significant unresolved emission near the X-ray position of \hess, which we characterised by fitting a point source in the image plane using the CASA IMFIT task. In the higher signal-to-noise image (5.5 GHz), the position of this source was RA (J2000) = 18:32:45.154 $\pm$ (0.009s or 0.13''), DEC (J2000) = -09:21:57.6 $\pm$ 2.6'', which is fully consistent with the {\it Chandra} position~\citep{eger16}. At 5.5 GHz, we obtained the flux density of $S_\mathrm{5.5}$ = 211$\pm$31 $\mu$Jy. At 9 GHz, $S_\mathrm{9}$ = 118$\pm$37 $\mu$Jy. Taking the conventional form of the spectral index $\alpha$ as in $S_\nu\sim\nu^\alpha$, one obtains $\alpha=-1.18^{+1.04}_{-0.88}$.

The steep spectrum does not suggest the radio source to be an AGN which should have a rather flat spectrum. Within the \gr~binary scenario, this radio counterpart may be due to the synchrotron radiation from shocked electrons or a putative radio pulsar.

\section{Swift/XRT observations of HESS~J1832$-$093}
Long-term X-ray monitoring is essential to check for any flare-like emission or variability of the source in X-rays. 
\citet{eger16} reported X-ray observations up to MJD~57291. We obtained 41 additional measurements from MJD 57460 to MJD 58287. The exposures ranges from $<1$~ksec to 5~ksec. Most of them are useful to build a long-term X-ray light curve, but the data qualities are not good enough for meaningful spectral analyses. In particular, the photon index ($\Gamma_\mathrm{X}$) and column density ($N_\mathrm{H}$) is not constrained by the Swift data, therefore we take the value of $N_\mathrm{H}$ to be $9.5\times10^{22}cm^{-2}$ and $\Gamma_\mathrm{X}=1.5$ which is obtained from a joint fit of the {\it Chandra}, {\it XMM-Newton}, and {\it NuSTAR} spectra~\citep{mori17}. This gives a count-to-energy flux ratio of one count to 1.382$\times$10$^{-10}$~ergs~$cm^{-2}~s^{-1}$ (2--10~keV).

To extract the XRT light curve, we used the \textit{Swift} online analysis tool\footnote{\url{http://www.swift.ac.uk/user_objects/}} \citep{2007A&A...469..379E,2009MNRAS.397.1177E} to take a good care of bad pixels, vignetting, and point spread function (PSF) corrections of the data. All parameters were left at program default values with the option \textit{binning by observation} chosen. 

\begin{figure}
    \epsscale{1.}
   \plotone{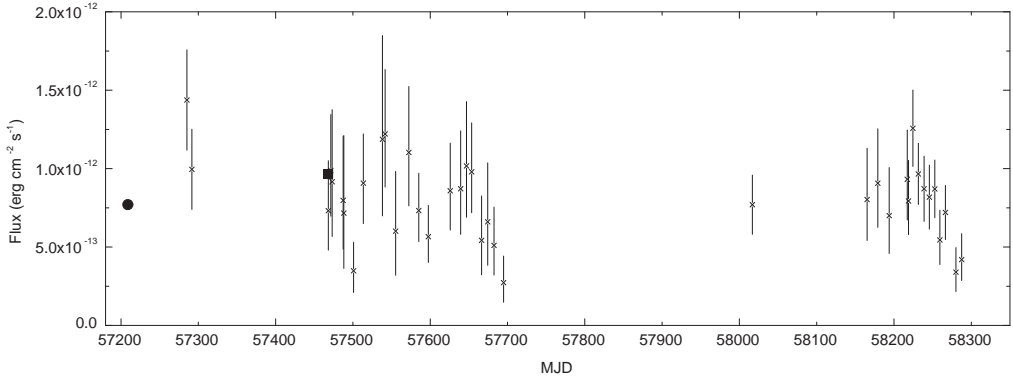}
      \caption{X-ray (2--10~keV) light curve (beyond MJD 57200) of HESS~J1832$-$093 obtained using Swift/XRT(asterisk), {\it Chandra} (circle), and {\it NuSTAR} (square).}
      \label{fig:swlc2}
   \end{figure}

The light curve (Fig.~\ref{fig:swlc2}, beyond MJD 57200) is shown together with the {\it Chandra} and {\it NuSTAR} data as presented in~\citet{mori17}. Fitting the full XRT count rate (including those before MJD 57200) with a constant flux returns a $\chi^2$(d.o.f.) value of 38.5(40). Therefore, no significant variability can be inferred solely from the Swift/XRT data due primarily to the large error bars. We also attempted to search for periodicity at time scales $>$0.5 day. No signifiant signal is seen, which again is subject to the low count rate and/or insufficient sampling seen by Swift/XRT. 

\section{discussion}

Among the five firmly identified TeV \gr~binaries so far, PSR~B1259$-$63, LS 5039, LS~I$+$61~303, and 1FGL~J1018.6$-$5856 all exhibit a peak in their SED at MeV-GeV energies, while HESS~J0632$+$057 may also have been detected in this energy band~\citep{hess0632_gev}. A MeV-GeV peak may be due to synchrotron emission, see e.g. PSR B1259-63 by~\citet{abdo11,Tam_1259_2015}. For HESS J0632+057, with its SED peak possibly in TeV, an inverse-Compton (IC) dominated model has been considered~\citep{hinton09,yi17,barkov18}. 

\begin{figure}
\centering
\includegraphics[scale=.70]{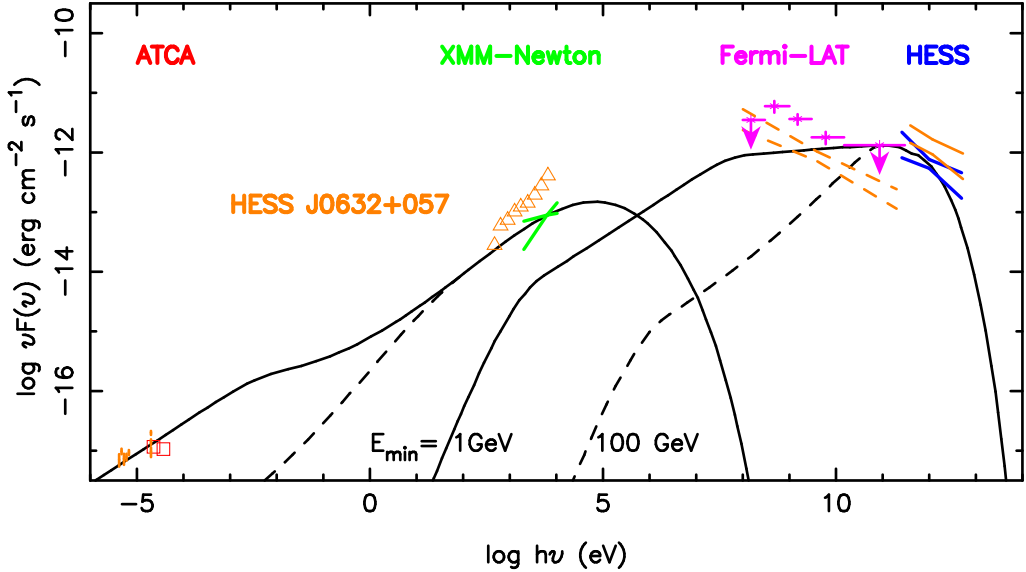}
      \caption{The SED of HESS~J1832$-$093 and {\bf the IC-dominated model used in \citet{hinton09} and \citet{eger16}. }
 The data points of HESS~J0632+057 \citep{skilton09} is also shown in orange for comparison. The dashed line is the {\it Fermi} spectrum of Fermi~J0632.6+0548, thought to be associated with HESS~J0632+057~\citep{hess0632_gev}.}
      \label{fig:sed}
\end{figure}

Based on our {\it Fermi} analysis results shown in Fig.~\ref{fig:sed}, {\bf the SED, i.e. the GeV band,} of \hess~{\bf looks like} neither the ones of former four binaries, nor the one of HESS~J0632$+$057. Assuming the GeV emission is indeed due to \hess, it would suggest that \hess~is spectrally different from those previous identified TeV \gr~binaries. {\bf We test two simplified one-zone models to explain its SED, a synchrotron-dominated model (not shown) and a IC-dominated model (shown as in Fig.~\ref{fig:sed}),} both of which face difficulties to explain the spectrum of 4FGL J1832.9-0913. {\bf Adding alternative sources can help explaining the GeV spectrum, and the SNR G22.7-0.2 can very well be such an alternative source. As discussed in Section.~\ref{sect:lat_analysis}, the possible extended morphology of 4FGL J1832.9-0913 also support this theory of alternative sources.}
If it is the case, the rather low break energy ($\sim$650~MeV in the PLE fit, see Sect.~\ref{sect:lat_analysis}) is similar to the cases of middle-aged SNRs, e.g., W44~\citep{abdo12_w44} and W51C~\citep{abdo09_w51c}. Future planned missions including eastrogam~\citep{eastrogam_white_paper} and AMEGO~\footnote{\url{https://asd.gsfc.nasa.gov/amego/}}, with their better PSF below 1 GeV, will certainly help. 

\section{Conclusion}

The X-ray and TeV spectrum of \hess~is similar to other \gr~binaries, e.g., HESS~J0632$+$057, as noted before, suggesting \hess~is also a \gr~binary. In this work, we present {\bf several additional observations to further explore this scenario:}

\begin{itemize}
\item We present the first NIR spectroscopy of the IR counterpart of \hess, which identifies it as an O-, or less likely, early B-type star; this result is consistent with a photometric measurement~\citep{mori17};
\item The rather steep ATCA spectrum does not suggest a radio AGN nature, and within the \gr~binary scenario, such a spectrum may be consistent with the synchrotron radiation from shocked electrons or a radio pulsar residing in the binary. Further radio observations with more accurate spectral measurements and dedicated radio pulsar timing search are needed to confirm the latter scenario.
\item {\bf The GeV source -- 4FGL J1832.9-0913 is} located just $\sim$10$\arcmin$ from \hess, and well within the radio emission of the SNR G22.7-0.2. There is no signifiant evidence for variability of this source. Evidence of extension at a radius of 12$\arcmin$ {\bf and the unique GeV spectrum which is difficult to be explained by conventional $\gamma$-ray binary models seem} to support the SNR scenario. In this case, \hess~remains one of the only \gr~binaries (candidates) without Fermi-LAT counterpart. Future missions with better PSF will certainly help for further verification of the nature of 4FGL J1832.9-0913. 
\item \citet{mori17} found a $\sim$50\% X-ray flux variability based on the {\it XMM-Newton} and {\it NuSTAR} observations. However, no significant variability can be inferred from the Swift/XRT data due primarily to the large error bars. This indicates that regular monitoring using sensitive X-ray instruments are needed to constrain the real X-ray light curve. Such a light curve is a possible way to identify the orbital period of \hess, as the case for HESS~J0632$+$057.
\end{itemize}

{\bf Ultimately, we find that the radio, IR, X-ray, and TeV point sources are associated with each other based on positional coincidence and they all support this \gr~binary hypothesis, while 4FGL J1832.9-0913 is probably due to alternative sources, i.e., SNR G22.7-0.2.} 

\acknowledgments
This research made use of data supplied by the High Energy Astrophysics Science Archive Research Center (HEASARC) at NASA's Goddard Space Flight Center, and the UK Swift Science Data Centre at the University of Leicester. We thank ATNF for the rapid scheduling of the ATCA observations. The Australia Telescope Compact Array is part of the Australia Telescope National Facility which is funded by the Australian Government for operation as a National Facility managed by CSIRO. This work is supported by the National Natural Science Foundation of China (NSFC) through grants 11633007, 11661161010, and U1731136.

\end{document}